\documentclass[12pt]{elsarticle}

\usepackage{natbib}
\usepackage{graphics}
\usepackage{graphicx}
\usepackage{epsfig}

\usepackage{amssymb}

\begin{document}

\begin{frontmatter}

\title{Generalization of Linearized Gouy-Chapman-Stern Model of Electric Double Layer for Nanostructured and Porous Electrodes: Deterministic and Stochastic Morphology}
\author{Rama Kant$^*$ and  Birla S. Maibam}
\address{Department of Chemistry,\\ 
University of Delhi,\\ 
Delhi 110007, India\\
Email address: rkant@chemistry.du.ac.in \\
URL: http://people.du.ac.in/$\sim$rkant}
\date{\today}


\begin{abstract}
We generalize linearized Gouy-Chapman-Stern theory of electric double layer for  nanostructured and morphologically disordered electrodes. Equation for capacitance is obtained using linear Gouy-Chapman (GC) or Debye-H$\rm{\ddot{u}}$ckel equation for potential near complex  electrode/electrolyte interface. The effect of surface morphology of an electrode on electric double layer (EDL) is obtained using ``multiple scattering formalism" in surface curvature. The result for capacitance is expressed in terms of the ratio of  Gouy screening length and the local principal radii of curvature of surface. We also include a contribution of compact layer, which is significant in overall prediction of capacitance.
Our general results are analyzed in details for two special morphologies of electrodes, i.e. ``nanoporous membrane" and ``forest of nanopillars". Variations of local shapes and global size variations due to residual randomness in morphology are accounted as curvature fluctuations over a reference shape element. Particularly, the theory shows that the presence of geometrical fluctuations in porous systems causes enhanced dependence of capacitance on mean pore sizes and suppresses the magnitude of capacitance. Theory emphasizes a strong influence of overall morphology and its disorder on capacitance. 
Finally, our predictions are in reasonable agreement with recent experimental measurements on supercapacitive mesoporous systems. 
  
\end{abstract}

\begin{keyword}
Gouy-Chapman-Stern theory, disordered electrodes, curvature fluctuations, Deterministic and Stochastic Morphology

\end{keyword}

\end{frontmatter}
\section{Introduction}

Electrochemical study on curved nanostructured surfaces (CNS) is due to the requirement to 
develop efficient energy generating and storage devices \cite{Simon}, electro-mechanical systems \cite{Baughman} and their applications in 
nanofluidics \cite{Schoch}. Recently, the electrochemical supercapacitive behavior of porous carbon materials\cite{Li}, e.g. carbide derived carbons (CDC) \cite{Simon,Gogotsi2003}, activated carbon (ACs), graphitic carbon, carbon nanotubes (CNTs)\cite{Baughman} etc., have acquired intense focus. These supercapacitive systems mainly based on the electrical double layer (EDL), requires surfaces with high specific area and volume with proper pore size and shape control for the efficient access of ions to obtain high energy storage as well as high power.   Hence, the electrical double layer (EDL) formed at the complex nanostructured or disordered  interface is a major focus for the research\cite{Schoch,Li,Ohkubo,Rangarajan80,Conway,Bazant2004, Bazant2009, Biesheuvel, Mustafa,Eikerling}.

Recently, Gogotsi and coworker were successful in developing supercapacitor with well controlled pore sizes in porous carbide derived 
carbon material (CDC) \cite{Simon, Gogotsi2003, Chmiola} which shows an  anomalous behavior in the capacitance\cite{Celine}.
These experimental results show three regimes in the capacitance vs pore size data\cite{Simon, Gogotsi2003, Chmiola}: (i) 
shows a nonlinear increase in capacitance, (ii) shows the transition from 
micropores to mesopores capacitance with a minimum, (iii) shows the anomalous increase and a maximum in capacitance of pores below 1 nm. 
Huang and coworker \cite{HuangJMR,HuangAC, Feng} proposed a heuristic model for the capacitance of such a problem with an assumption of cylindrical pores. They included the effect of finite pore sizes and proposed three models for three different pore size regimes: (i) micropore regime ($ < $ 2nm)- the electric wire-in-cylinder capacitors (EWCCs) model,\cite{HuangAC} (ii) 
mesopores regime ( 2-50 nm)-  the electric double cylinder capacitors (EDCCs) model \cite{HuangAC}    and for pores  in macropore regime ($>$ 
50 nm) where curvature is not longer significant, the parallel plate capacitor model was proposed. The capacitance (C) per unit area (A) of EDCCs model is given as\cite{HuangAC} 
\begin{equation}
 C/A = \frac{\epsilon_r \epsilon_0 }{ b\, \ln[b/(b-d)] }  \label{C1}
\end{equation}
 and for  EWCCs  is \cite{HuangAC}
\begin{equation}
C/A = \frac{ \epsilon_r \epsilon_0 }{ b\, \ln(b/a_0) }  \label{C2}
\end{equation}
 in which $A $ is the area of cylindrical pore, $\epsilon_r $ is the electrolyte dielectric constant, $\epsilon_0$ is the permittivity  of vacuum, $d$ is the effective thickness of the electric double layer (the Debye or Gouy length), $a_0$ is an effective size of the 
counter-ions in core and  $b$ radius of the outer cylinder. 
Computer simulations are becoming standard tools to study EDL on planar geometries\cite{NagyJPCB11,TorrieJPC82,BodaJCP02} but recently applied to curved  and porous systems\cite{KiyoharaJCP10,YangJACSJACS09,GoelJPCB11,ShimACSNano10}, with  
accounting ion size, ion-ion correlation\cite{TorrieJPC82,BodaJCP02} and ion-solvent interaction\cite{Tanimura2003, Tanimura2007}. Also simulations of EDL with accounting the dependency of the electrolyte dielectric permittivity on the local electric fields\cite{WangJPCC11} in sphere and  with
electrodes made of closely packed monodispersed mesoporous spheres\cite{WangAC11}. These simulations do not account for the 
general geometrical features and morphological fluctuation in the electrode. A general theory must account for the geometrical and morphological feature of an electrode that affects the capacitance.

The porous electrode materials are ubiquitously complex. 
The  modelling such systems is difficult, and complexity arise due to pore structure with inter connected three-dimensional connectivity of pores of nonuniform shape and size. The morphology of the pores and particulate materials may be idealized to various forms - prolate, oblate, ellipsoidal, spheroidal, tubular, etc.  These forms and morphologies can easily be identified with their local curvatures. But for the real pore space will be a combination of these idealised shapes or can be looked upon as fluctuations in pore structure around one of these idealized shapes hence identified with statistical property of their curvatures.  The electrochemical capacitance is strongly influenced by the morphology of nanoporous electrode materials \cite{Gryglewicz}. 
Hence, there is a need for systematic rigorous theory to understand the capacitance of nanostructured and disordered electrodes\cite{Yoon2010} which  interpret  capacitance behavior  of such electrodes within the frame work of electric double layer theories\cite{Dickinson,HuangJMR_diffuse}.

In this article, we develop (linearized) Gouy-Chapmann-Stern level theory for an arbitrary morphology.  Simplicity of the model is maintained through linearization while ignoring the microscopic interaction, viz, ion-solvent interaction and ion-ion correlation,  which may influence the ion- distribution. The theory is based on segregation of the compact layer and diffuse layer regions with an assumption of validity of the linearized Poisson-Boltzmann equation in the diffuse layer region while compact layer corrections are included at the level of local capacitance density. In section 2, we understand the effect of geometry and topology on the diffuse layer capacitance.
In section 3 we have included the compact layer contribution to diffuse layer capacitance on an arbitrary curved surface electrode. In section 4 we develop a theory for the nanostructured electrode with random morphology, viz, curvature fluctuation in a porous system arises due to intrapore roughness and interpore size variations. General results are applied for detailed analysis for the porous membrane and forest of nanorods electrode.
Comparison with recent experimental data of capacitance to pore size is presented in section 5. Finally, the summary and conclusions are discussed in section 6.

\begin{figure}[ht]
\includegraphics[scale = 0.5]{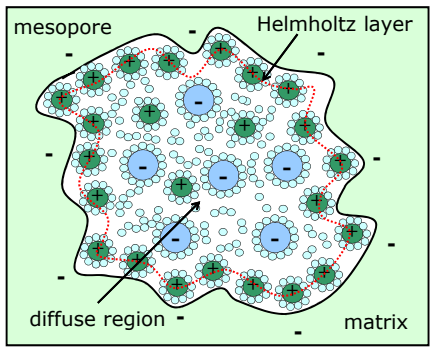}
\caption{Schematic model of EDL formed in arbitrary nanoporous electrode showing the Helmholtz layer and the diffuse layer.}
\end{figure}

\section{Model of Diffuse Layer near Curved Surfaces}

Historically, the concept of EDL is dated back to 1879.  It was Helmholtz \cite{Helmholtz}, who first conceived the idea that adsorption of opposite charge ions from solution on an electrode surface with excess or deficiency of charge results in a situation like a parallel plate capacitor of opposite charge with few nanometer thickness. The capacitance density of Helmholtz layer \cite{Conway} is
\begin{equation}
c_{H} = \frac{\epsilon_0 \epsilon_H}{r_H} \label{E1a}
\end{equation}
where $\epsilon _0$ is the dielectric constant of free space, $\epsilon _H$ is the HL relative dielectric constant, and 
$r_H$ the thickness up to the outer Helmholtz plane (OHP).  Nagy et al. showed that the dielectric constant in  HL  capacitance ($\epsilon_H$) should be the average of inner layer and diffuse layer dielectric constants due to the polarization charges induced on the boundary of the Helmholtz and the diffuse layer \cite{NagyJPCB11}.

Later Gouy\cite{Gouy} and Chapman\cite{Chapman} included the fact that the ions are mobile in the electrolyte solution due to thermal motions and 
developed a mathematical treatment  based on combined application of Boltzmann's energy distribution equation and Poisson's 
equation. However, the Gouy-Chapman model is known to over estimates of the EDL capacitance for planar electrode. The divergence was overcome by Stern\cite{Stern} in 1924 where he combined in series both the HL capacitance ($c_{H}$)  and the Gouy-Chapman capacitance of a planar electrode ($c_{\rm GC}$) as: 
$\frac{1}{c_{\rm dl}} = \frac{1}{c_H} + \frac{1}{c_{\rm GC} }$, where $c_{\rm GC} = \epsilon_{G}\kappa_{\rm eff}/4\pi$ and $ \kappa_{\rm eff} = \kappa \, \cosh(e \beta \phi/2)$. The Debye-H\"{u}ckel screening length,  $\kappa^{-1} = \ell_D = (\epsilon k_B T/4 \pi 
e^2 \Sigma n^{0}_{i} z_{i}^{2})^{1/2}$ where $T$ is temperature, $k_B$ the Boltzmann constant, $\epsilon$ the dielectric 
constant of bulk electrolyte, $n^{0}_{i}$ is the number density of i-th ion with charge $z_{i} e$ and e is the 
electronic charge.

Here we will like to develop a model for diffuse layer near a curved surface. Using Gauss 
law,  the local charge density ($ \sigma $) in electric double layer is: $\sigma = ({\epsilon}/{4\pi}) ({\partial\phi}/{\partial n})$; where $\phi$ is the  potential relative to the bulk solution and $\partial/\partial n\equiv \hat{n}\cdot\nabla$ is the  normal derivative (with unit normal vector ($\hat{n}$) to the surface).
The specific differential capacitance of a diffuse double layer is 
defined as $c_{G}= {d \sigma}/{d \phi_0}$, where $\phi_0 $ is the  potential difference applied at the interface. 
 Differentiating  $\sigma$ gives  capacitance density in terms of the electrostatic potential as 
\begin{equation}
c_{G} = \frac{\epsilon}{4\pi}\frac{d}{d \phi_0}\left( \frac{\partial\phi}{\partial n} \right) \label{E2}
\end{equation}
here $\epsilon$ is the dielectric constant of the bulk electrolyte and ${\partial \phi}/{\partial n} = \hat{n}\cdot\nabla \phi $  the 
inward normal derivative  of potential to the surface which is a functional of surface potential. 

The potential in Eq.~\ref{E2}  can be obtained using Poisson-Boltzmann equation (PBE) and Tessier and Slater\cite{Tessier} have shown the validity of PBE when the system size is reduced to 
nanoscopic dimensions.
Daikhin et al. \cite{Daikhin1996} showed that result for the nonlinear PB theory of capacitance at a rough electrode  
electrolyte interface is related to result for the linear version of PB Equation. On similar lines to include electrodes of arbitrary morphologies, we developed a theory based on simplified linearized Poisson-Boltzman equation. The effective nonlinear contribution of PB equation may 
 be included in the linear solution by replacing the inverse Debye length ($\kappa$) with a potential 
dependent inverse effective Debye or Gouy length $ \kappa_{\rm eff} = \kappa \, \cosh(e \phi/2k_B T)$ \cite{Lust2004,Daikhin}. 
The solution of Poisson's equation for the potential $\phi$ relative to the bulk electrolytes can be obtained under the condition of strong or weak potential approximations. 
Hence, the simplifying assumption used in our work is that the regions near an electrode are divided into a strong electric field region, i.e. the compact layer, and the region beyond the compact layer has a relatively weak electric field, i.e.  the diffuse layer (hence assumed to obey linearized Gouy-Chapman equation or Debye-H\"{u}ckel equation ).  

The linearized PBE  or the linearized Gouy-Chapman equation, for potential ($\phi(r)$) relative to the bulk solution, is written as 
\begin{equation}
(\nabla^2-\kappa^2)\phi(r) = 0. \label{GCE}
\end{equation}
where  the interfacial potential w.r.t. bulk is taken as constant and the bulk electrolyte potential is assumed to be 0.
The solution of the Helmholtz equation or Debye-H\"{u}ckel equation for 
arbitrary geometries and boundary conditions ({\it Dirichlet, Neumann or Robin}) are known in the form of  ``multiple 
scattering expansions" \cite{Balian, Duplantier90, Duplantier91, Duplantier1990,Kant_thesis}. Also `` multiple scattering expansion in curvature" results accounting for the arbitrary geometry of the surface is analysed for the diffusion problems\cite{Kant_thesis,Kant03, Kant97, Kant95,Kant94,Kant93}. Adapting a similar methodology to solve the electrostatic potential of electric double layer, we will obtain the capacitance of EDL near surfaces of arbitrary shapes in Debye-H\"{u}ckel or linearized Gouy-Chapman regime. The convergence of the screening length power series is applicable in a strong screening regime, viz, the Debye-H\"{u}ckel screening length, $\kappa^{-1}$ is smaller than any scale of radius of curvature. Through local and global curvatures brings in the geometric and topological features of surface are expected to play a role in the formation of the electric double layer near a curved surface. The expansions will be obtained in powers of $\kappa^{-1} /R$, where R is a typical radius of curvature.

 We use the method of Green function in order  to obtain the various order of scattering terms depending on surface curvature, 
the detailed calculation is shown in appendix.  Now the capacitance density at $\alpha$-th point of surface in terms of Green function is obtained using Eq.~\ref{E2} as 
\begin{eqnarray}
 c_{G}(\alpha) & = & \frac{\epsilon \kappa^2}{4\pi} \int_V d^3 r' \frac{\partial G(\alpha^{+} ,r')}{\partial n_\alpha}  
\nonumber\\ 
& = & \frac {\epsilon \kappa^{2}}{4 \pi}  {\int_V d^3 r' }\left[  2 \frac{\partial G_0(\alpha ,r' )}{\partial n_\alpha} \right.
\nonumber \\ 
&& \left. - 2^2\int \frac{\partial G_0(\alpha,\beta)}{\partial n_\alpha}\frac{\partial G_0(\beta,r')}{\partial n_\beta} dS_\beta \right. 
\nonumber\\ 
&& \left. + 2^{3} \int \frac {\partial G_0(\alpha , \beta)}{\partial n_\alpha} \frac{\partial G_0(\beta, \gamma)}{\partial n_\beta} \frac {\partial G_0(\gamma, r')}{\partial n_\gamma} dS_\beta dS_\gamma 
-  \cdots \right]
\end{eqnarray}
The local shape of the interface is given by mean ($H_{\alpha}$) and Gaussian curvatures ($K_{\alpha}$) at the $\alpha$-th point on the surface. 
Expressing the various orders of scattering terms are obtained in terms of two curvatures of the surface, the capacitance density of diffuse layer is (see appendix for details)
\begin{equation}%
c_G (H_{\alpha},K_{\alpha}) = \frac{\epsilon \kappa}{4\pi}\left[  1- \frac{1}{\kappa}H_{\alpha} - \frac{1}{2\kappa^2}(H^2_{\alpha}-K_{\alpha}) + \cdots\right]   \label{E17}
\end{equation}
where the local quantities $ H_{\alpha} $ and $ K_{\alpha} $ are defined as (dropping subscript for simplicity), $H = ({1}/{R_1 + {1}/R_2 })/2$ and $K = {1}/{R_1  R_2}$. Retaining Eq.~\ref{E17} upto second  
order in curvature expansion where two curvatures satisfy: $(H^2-K) = {1}/{4}\left( {1}/{R_1} - {1}/{R_2}\right)^{2}$ and $(H^2-K) \geq 0 $.   
Eq.~\ref{E17} clearly shows that the geometric dependence of the capacitance is controlled by second and third term through  $H$ and $K$. If $H$ = 0 and $K$ = 0, the diffuse layer capacitance simplifies to the flat surface ( for small applied potential) Gouy-Chapman capacitance \cite{Daikhin}.  
The local capacitance result for an arbitrary curved electrode has three terms. The first term is dependent on the solution properties, second term is  dependent on geometry (independent of ionic concentration)  but third term represents the coupling between geometrical and solution properties. 
Eq.~\ref{E17} for diffuse layer capacitance can be interpreted as an effective parallel curved surface capacitor, where two curved surfaces are separated by an effective distance $\kappa^{-1} /\left[  1- \frac{1}{\kappa}H_{\alpha} - \frac{1}{2\kappa^2}(H^2_{\alpha}-K_{\alpha})\right]$. The local separation between electrode and effective surface in solution depends on the composition as well the local curvature of the electrode.

\begin{figure}[ht]
\includegraphics[scale = 0.6]{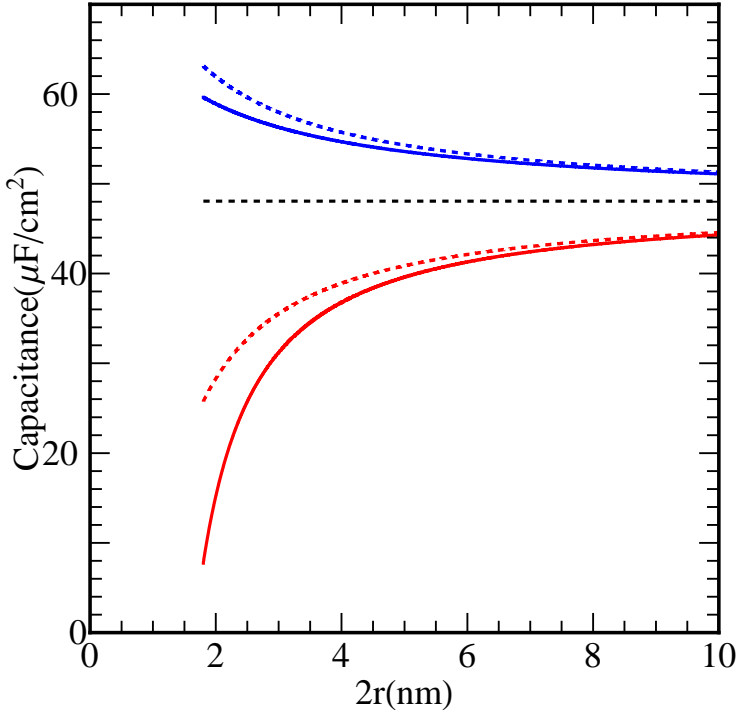}
\caption{Theoretical capacitance density vs diameter plots for cylindrical pore and rod electrodes.   The linearized Gouy-Chapman (LGC) capacitance curved surface (dotted lines) and attuned LGC capacitance (solid lines) after inclusion of Helmholtz layer size correction for pore (red line) and rod (blue line) geometries.  The plots are generated using $r_H$ = 0.35 nm, $\epsilon_H$ = 6 and $\epsilon$ = 38 for  an organic electrolyte at 298 K.}
\end{figure}

We illustrate use of Eq.~\ref{E17} through its application to a cylindrical pore and a rod geometries.  Fig. 2 shows the plot  for the capacitance of these simple geometries and their dependence on size (2 $r$). Two curvatures for a cylindrical pore, mean curvature $ H= 1/2r $ is constant and the Gaussian curvature  $K = 0$. Similarly, for a cylindrical rod, $ H= -1/2r $ while $K = 0$. In Fig 2 we show two cases of capacitance, viz. linearized Gouy-Chapman (LGC) model  capacitance for curved surfaces (dash lines)  and LGC capacitance after Helmholtz layer size correction we call this attuned LGC capacitance (AGC) for curved surfaces (solid lines), viz, for a cylindrical pore the attuned curvature is $ H= 1/(2(r-r_H ))$ and for a cylindrical rod the attuned curvature is $ H= -1/(2(r+r_H ))$. The LGC  and AGC capacitance curves for the pore have smaller values compared to flat surface. 
In the case of rod (red lines) we see that the LGC capacitance value increases with reduced rod diameter while AGC this rise  weakens due to larger attuned rod diameter. 
This finding for LGC is  similar to Compton and coworkers report of enhancement of capacitance in diffuse layer for hemispherical and cylindrical  electrode \cite{Dickinson, Henstridge}. 
But with  increasing sizes, all electrodes reach curvature independent capacitance value and merge with the planar capacitance limit.  The point particle nature of electrolytes beyond the compact layer, allows it to be extended up to pore size equal to mono-layer of ions or say upto the compact layer size.  In short, it is important to emphasize that our theory will be applicable to a porous system for pore sizes $2r>2r_H+l_D$, i.e. sufficient to accommodate compact as well as the diffuse layer inside them.
Extension of curves below the pore diameter $2r_H+l_D$ is allowed due to point particle nature in diffused layer, but in reality we will require contributions from volume exclusion, material space charge corrections etc.  for an accurate prediction of capacitance.
Here we show the emergence of morphological influence on capacitance is mainly due to  diffuse layer. It will be significantly seen in cases when the magnitude of the smallest radius of curvature (R) follows: $r_H < R \le (2r_H+l_D)$. 
   The existence of  minimum in the capacitance of highly confining pores is reported in literatures\cite{Simon,Chmiola,Celine,HuangAC} but its explanation will require inclusion of material capacitance of the electrode. The large radius of curvature limit of capacitance of pore as well as  rod has curvature independent capacitance regime (as of planar surface). The experimental observations \cite{Celine, ChmiolaAC} of maximum capacitance at a critical pore diameter which corresponds to the effective thickness of HL may require  inclusion of space charge contribution form the electrode material in our theory.

The diffuse layer capacitance given by Eq. (\ref{E17}) may be represented for simple curved  surfaces as 
\begin{equation}
c_{G} = c_{0} \left( 1 + a ~ \frac{l_{D}}{r_{a}} + b~ \frac{l_{D}^{2}}{r_{a}^{2}} \right) \label{SE1}
\end{equation}
where $c_0 = {\epsilon_0 \epsilon_D}/{l_D}$ is planar capacitance density, $(a,b)$ are pairs of dimensionless numbers depending on the shape of a nanoelectrode, viz.  (0,0) for a plane, (-1/2, -1/8) for a tube, (1/2, -1/8) for a rod, (-1,0) for a cavity and (1,0) for a sphere. Effective attuned radius due to compact layer ($r_{a}=r\pm r_H$) on curved surfaces, where $+$ and $-$ signs are assigned to convex (e.g. sphere and rod) and concave (e.g. tube and cavity) surfaces, respectively.  

The diffuse layer capacitance obtained in Eq.~\ref{SE1} can be looked upon as the capacitance between two curved surfaces, e.g. electrode surface and another effective surface in solution with same symmetry at distance $l_D/(1 + a ~ \frac{l_{D}}{r_{a}} + b~ \frac{l_{D}^{2}}{r_{a}^{2}})$ (in the normal direction). More specifically, the capacitance of a cylindrical pore in Eq.~\ref{SE1} can be approximated 
as an effective concentric cylinder capacitor with their effective separation $l_D/(1 - \frac{l_{D}}{2r_{a}} - \frac{l_{D}^{2}}{8r_{a}^{2}})$ and for a  cylindrical rod electrode in presence of diffuse layer  is equivalent to wire in a cylinder capacitor with their separation $l_D/(1 + \frac{l_{D}}{2r_{a}} - \frac{l_{D}^{2}}{8r_{a}^{2}})$.

Another important quantity of physical interest is the total diffuse layer capacitance of a nanostructured electrode, $C_D$.  This is obtained by integrating the local diffuse layer capacitance density $c_G$ in Eq.~\ref{E17}, with respect to the interfacial surface as
\begin{equation}
C_D=\int  dS\, c_G= \frac{\epsilon \kappa}{4\pi}\left[ A - \frac{1}{\kappa}\overline{H} - \frac{1}{2\kappa^2}( \overline{H^2} -\overline{K}) + \cdots\right] \label{cdsa}
\end{equation}
$K$ gives the intrinsic property and  topological information for a surface through theorem of Gauss-Bonnet \cite{Carmo} as 
$\overline{K} = \int_S dS/(R_1 R_2) = 4\pi\left(N_c-N_h \right)$ where $N_c$ is number of connected surface and $N_h$ is the 
number of handles (genus or holes) in the surface. $H$ gives the extrinsic property and knowledge of space.  The morphological  
quantities are the geometric area $ A = \int_S dS$, integral of mean curvature $\overline{H} = \int_S H dS$, integral of square of mean curvature $\overline{H^2} = \int_S H^2 dS$ and  topological quantity,  integral of Gaussian curvature $\overline{K} = 
\int_S K dS$ can be calculated both for regular and random geometries.

\section{Inclusion of Compact Layer Contributions in Local  Diffuse Layer Capacitance }

The  capacitance of the EDL at Gouy, Chapman and Stern level is usually obtained using an analogy of capacitors in series.
Here we assume validity of  ansatz of combining two components of capacitances in series.  Hence local capacitance density ($c_{s}$) for curved interface is the local compact ($c_H$) and local diffuse layer ($c_G$) capacitance density in series  and written as: 
\begin{equation}
\frac{1}{c_{s}} = \frac{1}{c_H} + \frac{1}{c_G } \label{E1c}
\end{equation}
 Two limiting situations for the local capacitance are observed due to different functional dependence of the charge on the electrolyte concentration  and electrode potential. 
  
Two components of capacities are estimated in our analysis by dividing the regions near an electrode into a strong electric field region- the capacitance estimated using the Helmholtz formula. The region beyond the compact layer is the diffuse layer and has a relatively weak electric field. The diffuse layer capacitance is estimated through  AGC model for the curved interface. This approximation provides a workable simplicity to our model.  
 Combining Eqs.~\ref{E17} and \ref{E1c} we obtained the attuned (linearized) Gouy-Chapman-Stern (AGCS) capacitance density for an arbitrary geometry electrode as a function of mean and Gaussian curvatures:
\begin{equation}
c_{dl} (H,K)=\frac{ c_H c_0 \left[  1 - H l_D - (H^2 - K )(l_{D}^{2}/{2})\right] }{c_H + c_0 \left[  1 - H l_D - (H^2 - K )(l_{D}^{2}/{2})\right] } \label{E18}
\end{equation}
where  $c_0={\epsilon \kappa}/{4\pi}$ and $c_H = {\epsilon_H \epsilon_0}\slash {r_H}$, where $\epsilon_H $ is the  electrolyte dielectric constant in Helmholtz layer (HL) and $\epsilon_0$ is the permittivity of vacuum, $r_H$ is the  {\it effective thickness} of Helmholtz or compact layer. Here $H$ and $K$ are obtained at the Helmholtz surface (by local shifting of electrode surface in normal direction by distance $r_H$), respectively. The diffuse layer starts beyond the compact layer hence the curvature attuned at the Helmholtz layer is used in our calculation.
The effective or attuned radii of curvatures at the diffused layer are accounted by subtracting or adding $r_H$ from radii of curvatures depending on local convexity of surface. Combining compact layer capacitance with AGC capacitance partially removes point particle limitations as well as the divergence under $R/l_D \rightarrow 0$ limit. The local AGCS capacitance in Eq.~\ref{E18} has three terms in the square brackets. The first term in square bracket is dependent on the solution properties, second term is  dependent on geometry through the local mean curvature  but third term represents the coupling between geometrical and solution properties. Third term has mixed mean and Gaussian curvatures, viz $H^2 - K$, which is equal to 0 for sphere hence measure of deviation from local asphericity and follows constraint: $H^2 - K\ge 0$. Hence local coupling between geometry and ionic solution emerges due to deviation in local asphericity of surface.
 
\section{Capacitance for Random Morphology Electrodes}

Randomness arises due to disorder in porous material can be of three types: (1) pore roughness can be looked upon as intra-pore shape fluctuations  along the contour of the pore,  (2)  inter-pore size 
fluctuations over different pores and (3) fluctuations in length of pores. First two contributions of randomness can be accounted through fluctuations in mean curvature. Some examples are: carbide derived carbons (CDCs) have intrapore  fluctuations, template carbon materials (TCM) may have both intrapore fluctuations and interpore pore-size fluctuations,  activated carbons (ACs) and hierarchical porous graphitic 
carbon material (HPGC)\cite{Wang} may have intra-pore curvature fluctuations, pore size fluctuations and pore length 
fluctuations. Fluctuations in length of pore affects, mostly the dynamic response of pores \cite{Wang,Lust2004}. This case is 
not required in calculations as one is not looking in the dynamic aspect of this problem.

 The most appropriate method for characterizing the complex disordered structure are based on the statistical approach.  
Depending upon the nature of disorder in nanostructured or porous electrode one can have various distributions  of curvature.  We assume in our model of an electrode with small random surface roughness around its reference geometry and  is characterized in terms of an ensemble averaged mean and Gaussian curvatures, viz, $\langle H \rangle$ and $\langle K \rangle$.  The surface roughness is characterized in terms of their deviations from reference curvature: $H - \langle H \rangle$ and $K - \langle K \rangle$ with an ensemble average mean values:  $\langle H - \langle H \rangle\rangle=0$ and $\langle K - \langle K \rangle\rangle=0$ but finite variance $\langle (H - \langle H \rangle)^2\rangle$ (fourth order term in principle curvature, $\langle (K - \langle K \rangle)^2\rangle$, is assumed to be negligible).   In order to calculate the capacitance of disordered systems,  we take the statistical ensemble of various configurations in curvature space (for which the curvatures
are distributed).  Now the ensemble averaged capacitance density of diffuse layer is obtained from Eq.~\ref{E17} as
\begin{equation}
\langle c_G  \rangle = c_0 \left[  1-  l_D \langle H \rangle -\frac{l_{D}^2 }{2}( \langle H^2 \rangle -   \langle K \rangle) + \cdots\right]   \label{E17A}
\end{equation}
where $\langle\cdot\rangle$ represent an ensemble average over all possible random surface configurations and $l_D=\kappa^{-1}$. Eq.~\ref{E17A} has three terms: first term is dependent on only solution ionic concentration, second term is purely dependant on ensemble averaged mean curvature while third term is due to coupling between ionic solution and electrode morphology.   The mean capacitance density $\langle c_G  \rangle$ of diffuse EDL will be strongly affected whenever  the radii of curvature will be comparable to diffuse layer thickness, which is a strong function of electrolytes concentration. 

Deviation in an ensemble average capacitance density is proportional to the difference of the curvature of surface and ensemble average curvature. The capacitance deviation of diffuse layer  ($\delta c_G =  c_G - \langle c_G \rangle$) is 
\begin{equation}
\delta c_G = c_0 \left[ l_D \left( H - \langle H \rangle \right) - (l_{D}/2) \left( H^2 - \langle H^2 \rangle \right) +  (l_{D}^2/2) \left( K - \langle K \rangle \right) \right] \label{SD1}
\end{equation}
The average deviation of capacitance is zero, i.e. $\langle \delta c_G \rangle = 0 $.  The mean square deviation of difference of capacitance density is obtained from Eq.~{\ref{SD1}} as
\begin{equation}
\langle \delta c_{G}^2\rangle = c_{0}^2 l_{D}^2 \langle (H -  \langle H\rangle)^2 \rangle + O(H^4) \label{SD2}
\end{equation}
An important feature of practical surfaces is the presence of small surface roughness characterized as small fluctuations in surface curvatures about its reference geometry. We have truncated higher-order curvature contributions in Eq.~\ref{SD2}.	
 It is obvious from Eq.~\ref{SD2} that the ensemble average of square of capacitance deviation is directly proportional to the ensemble  
average of square of deviation in mean curvature. Rewriting Eq.~\ref{SD2} in terms of a coefficient of morphological fluctuations or relative variance of the mean curvature, $\gamma^2 = 
\langle (H -  \langle H\rangle)^2 \rangle/\langle H\rangle^2$, we have 
\begin{equation}
\langle \delta c_{G}^2\rangle = \left(\frac{\epsilon}{4\pi}\right)^2 \gamma^2 \langle H\rangle^2 + O(H^4) \label{SD3}
\end{equation}
where $\gamma$ is a coefficient  of morphological fluctuations and  is a measure of relative deviation from the mean curvature. This variance in capacitance density is independent of concentration of electrolyte. Coefficient $\gamma < 1$ for weakly fluctuating interface and $\gamma \gg 1$ for strongly fluctuating interface. 

As mentioned before porous materials have complex spatial structures and are characterized by using morphological measures, viz, the mean (microscopic) geometric area of an electrode is $A$, ensemble average mean curvature  $\langle H \rangle $, ensemble average square of mean curvature $\langle H^{2}\rangle $, and the ensemble average of Gaussian curvature $\langle K \rangle$. For random ergodic fields\cite{Alder} which are statistically 
homogeneous over various configurations,  the (large) surface average over morphological quantities, viz, $\overline{H}/A$, $\overline{H^2}/A$ and $\overline{K}/A$ (as defined earlier),  can be related to the 
ensemble average morphological quantities as  $\langle H\rangle \equiv \lim_{A\rightarrow \infty}\overline{H}/A $, $\langle H^2\rangle\equiv \lim_{A\rightarrow \infty}\overline{H^2}/A $ 
and $\langle K \rangle \equiv \lim_{A\rightarrow \infty}\overline{K}/A $. 

These morphological measures are in general useful to characterize the structure of various materials, {\it e.g.} 
foams, gels, membranes, granular and porous electrode systems. These measures can be calculated for 
both deterministic and stochastic geometries, and are related through integral geometry  to {\it Minkowski functionals} 
\cite{Mecke}. The ensemble average morphological measures now can be related through above integral of mean curvature, 
integral of mean square curvature and integral of Gaussian curvature.  Hence for a large nanostructured surface, Eqs.~\ref{cdsa} and \ref{E17A} are related through: $\langle c_G \rangle \equiv\lim_{A\rightarrow \infty} C_D/A$. Since the reciprocal of compact and diffuse layers local capacitance is  additive in nature we will perform averaging over an ensemble of random configurations. The ensemble averaged (inverse) capacity is now
\begin{eqnarray}
 \langle 1/ c \rangle & = & \langle 1/ c_H \rangle + \langle 1/ c_G \rangle
\nonumber\\  
& = & 1/ c_H  + \langle 1/(\langle c_G \rangle +  \delta c_G ) \rangle
\end{eqnarray}
Using binomial expansion of relative deviation in capacity, its simplifies to following equation
\begin{eqnarray}
\langle 1/ c \rangle &  = & 1/ c_H  + 1/\langle c_G \rangle +   \langle \delta c_G^2\rangle /\langle c_G \rangle^3 + O( \langle \delta c_G^4 \rangle)
\label{E19A}
\end{eqnarray}
Eq.~\ref{E19A} for random morphology electrode can be written as the ensemble average of inverse of capacitance  as:
\begin{equation}
c_d=\left\langle \frac{1}{c} \right\rangle^{-1}   =  \left(\frac{1}{\langle  c_S \rangle} +\frac{\beta^2}{\langle c_G \rangle}\right)^{-1}\label{cap}
\end{equation} 
where $1/\langle  c_S \rangle = 1/ c_H  + 1/\langle c_G \rangle $ and  $\langle  c_S \rangle$ is mean compact and diffuse layer capacitance density for average surface morphology.  The relative mean square fluctuation in capacitance ($\beta^2=\langle \delta c_G^2 \rangle/\langle c_G \rangle^2$) is: 
\begin{equation}
\beta^2=\gamma^2\, l_{D}^2\,  (\overline{H}/A)^2/\left[  1 -  (\overline{H}/A) l_D - \left( (\overline{H^2}/A) - (\overline{K}/A)\right)(l_{D}^{2}/{2})\right]^2
\end{equation} 
where $\gamma$ represent the extent of fluctuations in surface morphology.

Using  an assumption of ergodic field \cite{Alder}, Eq.~\ref{cap} in combination with Eqs.~\ref{E17A} and \ref{SD3} simplifies to the mean  EDL capacitance  in two components. The non-fluctuating contribution ($\langle c_S\rangle$) from this equation is represented as
\begin{equation}
\langle c_S\rangle = (1/ c_H  + 1/\langle c_G \rangle)^{-1} =  \frac{ c_H c_0  \left[  1 -  (\overline{H}/A) l_D - \left( (\overline{H^2}/A) - (\overline{K}/A)\right)(l_{D}^{2}/{2})\right] }{c_H + c_0 \left[  1 -  (\overline{H}/A) l_D - \left( (\overline{H^2}/A) - (\overline{K}/A)\right)(l_{D}^{2}/{2})\right] } \label{E19}
\end{equation} 
The contribution from fluctuating component (${\beta^2}/{\langle c_G \rangle}$) in the capacitance which arise due to stochastic nature of morphology is
\begin{equation}
\frac{\beta^2}{\langle c_G \rangle}  =  \frac{1}{c_{0}}\; \frac{\gamma^2\; l_{D}^2 (\overline{H}/A)^2}{\left[  1 -  (\overline{H}/A) l_D - \left( (\overline{H^2}/A) - (\overline{K}/A)\right)(l_{D}^{2}/{2})\right]^3}\label{rancap}
\end{equation} 
Eq.~\ref{cap} in combination with Eqs.~\ref{E19} and \ref{rancap} is generalized average capacitance density ($c_d$) for an arbitrary random geometry. This expression relates the capacitance to 
the surface to the averaged mean and Gaussian curvatures of the random morphology electrodes. 
Hence, the capacitance density ($c_d$) of the complex morphology electrode is dependent on an overall surface morphological and topological characteristics, viz, $\overline{H}$, $\overline{H^2}$ and $\overline{K}$.

\subsection{Porous Membrane with Fluctuating Pore Sizes}

The complex nature of the nanoporous electrode may results in various distributions in surface shapes (curvatures)  and pore lengths. In this section, we model the influence of intrapore and interpores size fluctuations. Interpore size fluctuations arise due to the presence of polydispersity in pore sizes, which is often much larger than intrapore fluctuations. Hence, the nature of pore size and shape fluctuations is critical in determining capacitance of porous electrodes.  For narrowly distributed pore sizes, e.g., carbide derived carbons (CDC) micropores with the mean pore size is less than 2 nm, only the intrapore local surface fluctuations are significant. There can be low intra-pore fluctuations  possible without choking pore in the electrode. For broadly distributed mesopores not only the fluctuation in interpores sizes is important but also the intrapore  curvature fluctuations and hence there could be simultaneous contributions from both. The porous materials have pores of various sizes and shapes, e.g. activated carbons,  such systems are approximated as membranes electrode with micropores and mesopores.

 Here,  we confine our analysis to simple case of nanoporous materials as a porous membrane model  which has an array of cylindrical pores. These pores are allowed to fluctuate weakly, so that the topology of the membrane electrode is not changed.  Typical geometrical features of the membrane are characterized by pore separation distance ($w$), membrane thickness (pore length) ($l$),  the outer macroscopic geometric area  of membrane electrode ($2 A_0$) and number of pores ($N_h$). In order to calculate $N_h$, we assume a  simple model of membrane with a hexagonal array of cylindrical pores whose number of pore is given by $N_h$  = $2A_0/(\sqrt{3}(2r+w)^2)$\cite{Elliott}, where  $r$ is the radius of cylindrical pore. Since the number of pore per unit area is usually very large, in orders of $10^{12}$, the area of two planar surfaces, $2(A_0-\pi r^2 N_h)$ , connecting two ends of pores of the membrane is small and hence can be neglected. The major part of area contribution comes from the area inside pores.  The total area of such a porous surface is given by area of individual pore times pore density and, which is $A$ $\approx$ 2$\pi r l$ $N_h$.  For a cylindrical pore, mean curvature $ H= 1/2r $ is constant and the average mean curvature is  $\overline{H}/A = \int{H} dS/A = 1/ 2 r $. Similarly, square of mean curvature is: $H^2 = 1/ 4r^2$  and the average square of mean curvature  is: $\overline{H^{2}}/A = \int H^{2} dS/A = 1/ 4 r^2 $. The  average Gaussian curvature of a porous membrane is calculated using the theorem of Gauss-Bonnet as  $\overline{K}/A = \int K dS /A = 4 \pi ( N_c - N_h)$, where $N_c$ is number of connected surfaces. For a single membrane electrode, the connected component $N_c = 1$ and $\overline{K}/A  =  (2/r l )(1-N_h)/N_h \approx -2/r l$(for large $N_h$ electrode). 

Knowledge of various morphological measures for a given model allows us to illustrate the use of  Eq.~\ref{cap} or Eq.~\ref{E19} under smooth surface  limit $\gamma\rightarrow 0$ hence taking ion with the {\it effective ion size} in the HL or thickness of HL, $r_H$ and the diffused layer of the EDL is assumed to start after the Helmholtz layer. The effective or attuned pore radii for diffused layer are accounted by subtracting HL thickness $r_H$ from pore radius $r$, i.e. $r - r_H$.  Substituting the morphological quantities in  Eq.~\ref{E19}, an explicit equation for the smooth pore membrane is obtained as 
\begin{equation}
\langle c_S\rangle =  \frac{ c_H c_0  \left[  1 -   l_D/2(r-r_H) - l_{D}^{2}/8(r-r_H)^2 -  l_{D}^{2}/l (r-r_H) + l_{D}^{2}/N_h l (r-r_H)  \right] }{c_H + c_0 \left[  1 -   l_D/2(r-r_H) - l_{D}^{2}/8(r-r_H)^2 -  l_{D}^{2}/l (r-r_H) + l_{D}^{2}/N_h l (r-r_H)  \right]  } \label{E21}
\end{equation} 
Similarly, the relative mean square fluctuation in capacitance ($\beta^2$) due to curvature fluctuations in pore sizes or roughness along the contour of pores is
\begin{equation}
\frac{\beta^2}{\langle c_G \rangle}   =  \frac{1}{c_{0}}\; \frac{\gamma^2\; l_{D}^{2}/4(r-r_H)^2}{\left[  1 -   l_D/2(r-r_H) - l_{D}^{2}/8(r-r_H)^2 - l_{D}^{2}/l (r-r_H) + l_{D}^{2}/N_h l (r-r_H)  \right]^3}\label{ranpore}
\end{equation} 

 It may be noted from Eqs.~\ref{E21} and \ref{ranpore} for two components of capacitance density equation, $c_d=\left( \frac{1}{\langle c_S\rangle} +\frac{\beta^2}{\langle c_G \rangle}\right)^{-1}$, has a contribution from the Helmholtz layer, the diffuse layer and morphological parameters, viz, pore radius ($r$), pore length ($l$), number of pores ($N_h$) and coefficient of fluctuation  ($\gamma$). The capacitance of porous surfaces has a form of rational polynomial, viz, Pad\'{e} approximant \cite{Kant90}, in inverse attuned pore size $(r-r_H)$.  Eq.~\ref{E21} shows that the ionic concentration dependent Debye length $ l_D$  and attuned pore size $(r-r_H)$ are length scales that affect the capacitance of the porous membrane. 
As mentioned, earlier that the diffuse layer contributions in Eqs~\ref{E21} and \ref{ranpore} have solution property dependent term, geometry dependent term and a coupling term between solution and geometry. For pore size  $r-r_H =1nm$,  the relative importance of these three terms for $l_D\approx$ 1 nm (for 0.1M electrolyte solution) and is 1:1/2:1/8 and for $l_D\approx$ 3 nm (for 0.01M electrolyte solution) its 1:3/2:9/8. 
  Hence, we conclude that diffuse EDL will have a significant influence on the capacitance of a porous system.
 
In Figs. 3(a) and 3(b), we analyse the effect of Helmholtz layer thickness and the effect of change of concentration (or Gouy length) in the absence of the fluctuations in contour of the pore ($\gamma\rightarrow 0$), respectively. 
\begin{figure}[ht]
\includegraphics[scale = 0.5]{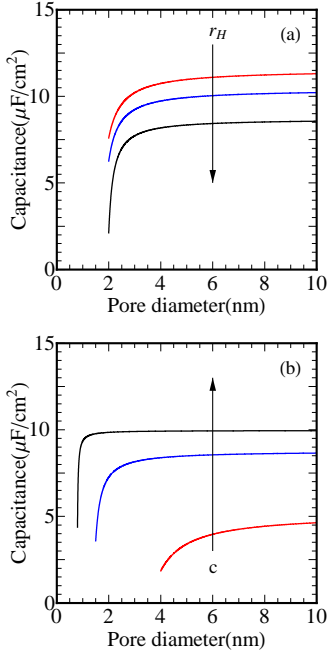}
\caption{ 
Capacitance plots for a porous membrane electrode with cylindrical pores obtained from Eq.~\ref{E21} for 1:1 organic electrolyte  with $\epsilon$ = 38  at 298 K. 
(a) Effect of compact layer thickness, $r_H$, on the capacitance profile. The plots are generated using concentration, c = 1 M, $\epsilon_H$ = 6 and $r_H$ = 0.35, 0.4 and 0.5 nm.  (b) Effect of concentration on capacitance. The plots are generated using $r_{\rm H}$ = 0.35 nm, $\epsilon_H$ = 6 while the concentration used in various curves are: 0.01, 0.1 and 1 M.}
\end{figure}
Fig. 3(a) shows the attuned (linearized) Gouy-Chapman-Stern (AGCS) capacitance density plots for the porous membrane electrode, but without pore diameter fluctuations ($\gamma=0$). The curves show the effect of HL-thickness  on overall capacitance. These plots are obtained using Eq.~\ref{E21} which  show two regions in capacitance behaviour and a crossover between them.  The point particle nature of ions in diffused layer allows us to extrapolate our capacitance vs pore-size plot in the narrow pore size regime, but this continuum model for the electrical double layer capacitance is valid upto pore size $2r_H+\ell_D$ hence prediction below this value will require dominant contribution from the electronic capacitance of the electrode. These observations usually indicated in several experiments but not captured by heuristic models\cite{Simon, Chmiola, HuangAC}. 
The  capacitance becomes independent of the pore size beyond 2 nm, while the experiments show dependence up to 10 nm size. This could be due to not inclusion of  roughness  or morphological fluctuations in porous electrodes.  The quantitative predictions near the highest capacitance region may require electronic space charge correction from the electrode as well as other contributions like adsorption, excluded volume, etc. in the high field compact layer region.
	
Fig. 3(b) shows the effect of concentration  and pore sizes on the GCS capacitance density. Three curves are obtained for different concentration of electrolytes (or $l_D$). There is a gradual disappearance and a simultaneous shift in the location of the minimum point with the decrease in concentration. Other conclusions can be drawn from these curves are: (a)  high concentration result, $\kappa r >1$, has a minimum value of capacitance at pore size  $2r = 2r_H +  \kappa ^{-1}$,  (b) low concentration results, $ \kappa r < 1$,  there is gradual reduction of capacitance. 
Anomalous increase in capacitance is observed in CDC where pore size is less than 1 nm \cite{Chmiola} but this cannot be explained purely on the basis of ionic capacitance.  
The capacitance has a sudden rising  region followed by a gently rising with increasing pore size and a crossover between these two regions is seen here. This is an interesting  finding for system with unimodal pore size and influence of the fluctuation in pore sizes in capacitance behavior will be shown in Fig. 4.

\begin{figure}
\includegraphics[scale = 0.7]{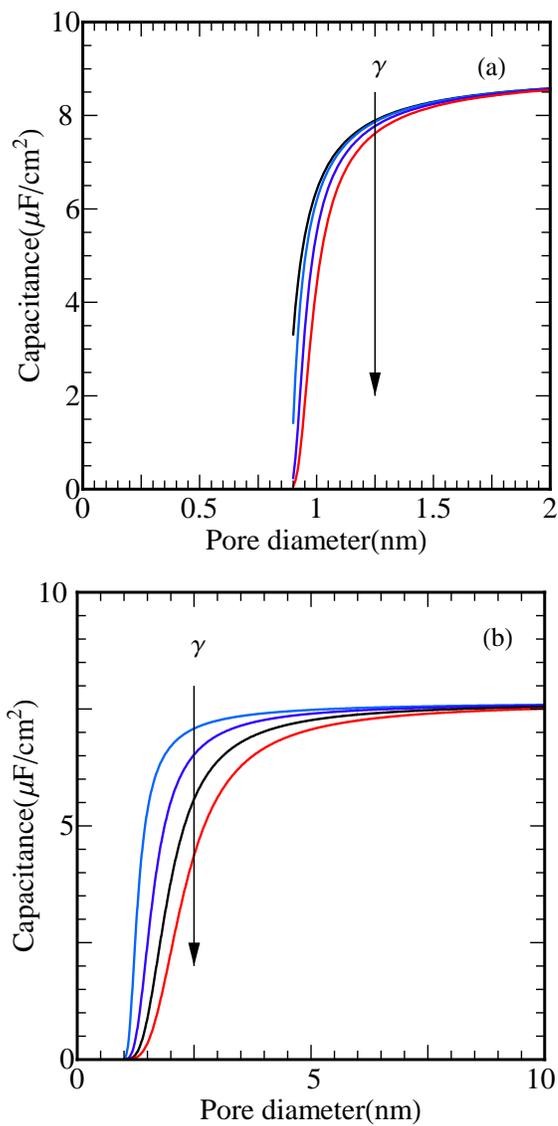}
\caption{Effect of the morphological fluctuations in a porous membrane system on capacitance vs mean pore size plots.
(a) Small fluctuation in pores sizes with $\gamma$ = 0, 0.3, 0.6 and 0.9. 
(b) Large fluctuation in pores sizes with $\gamma$ = 2, 4, 6 and 8. All plots are generated using $r_H$ = 0.35 nm, $\epsilon_H$ = 6, for 1:1  organic electrolyte  with $\epsilon$ = 38  at 298 K.}
\end{figure}
 
In order to see the effect of morphological fluctuation on capacitance density, we plot capacitance of a porous membrane electrode. The pore shape and size can fluctuate along the contour of the pore. Fig.~4 shows the effect of surface morphological fluctuations on the capacitance of a porous membrane electrode. Fig. 4(a) describes capacitance vs mean pore size in the presence of small fluctuations in the micropore regime. We observe from the curves that as we increase the value of surface morphological fluctuation parameter $\gamma$, the sudden rise region becomes weaker while the slow rise region becomes slightly faster rise.  All the curves merge around 2 nm and no effect of small fluctuation is seen beyond this. Thus, the small fluctuations in micropores are important upto 2 nm. Fig. 4(b) shows the effect on capacitance due to large fluctuations in mesopores regime. As we increase $\gamma$, the value of capacitance is lowered,  compare to smooth porous electrode. The effect of  large fluctuation is seen for pore diameter much larger  than 2$r_H$+ $l_D$. Increasing the mean pore size in mesopore regime reduces the influence of fluctuation, and finally, all the curves merge around 10 nm. No curvature effect is seen beyond 10 nm, which suggest that large morphological fluctuation enhance capacitance dependence on mean pore size upto 10 nm. This is in agreement with several experimental reports of capacitance vs pore size data (see comparison with experimental data). The small fluctuation ($\gamma$ $<$ 2) affects the micropore regime while for large fluctuations ($\gamma$ $>$ 2), affects the capacitance in the mesopore regime. For pore diameter less than 2 nm, the slope of the rapid rise region is almost same. In case of mesopores the fluctuation results in a decrease of capacitance. Larger is the fluctuation, larger is the pore size up to which the effect is seen.  The capacitance increases rapidly in the transition region between micro and mesopores followed by a gradual increase in  mesoporous regime and limiting plateau value of a flat electrode is observed at large mesopores. Hence large morphological fluctuations play an important role to enhance the capacitance density dependence on mean pore size.

\subsection{Forest of Nanorods with Morphological Fluctuations}

In this section, we obtain capacitance for the forest of nanorods and compare them to that of nanopores. These two systems differ only in convexity of their electrode surface. The  typical dimension of the forest of nanorods electrode is characterized by rod separation distance, ($w$), electrode thickness (rod length), ($l$),  the geometric area ($A_0$) and number of rods ($N_r$). In order to calculate $N_r$ we assume an electrode with a hexagonal array of cylindrical rods whose number of rods is given by $N_r$  = $2A_0/(\sqrt{3}(2r+w)^2)$, where  $r$ is the radius of cylindrical rod. The total area of such a forest of nanorods structured surface is given by area of individual rod times rod density ($N_r$) and it is $A$ $\approx$ 2$\pi r l$ $N_r$.
Our theory is applied to nanoforest electrode with inter rod separation ($w$) which can accommodate compact as well as a diffused layer between them, viz,  $w\gg 2r_H+l_D$. The capacitance for array of cylindrical rods is obtained by replacing,
$\overline{H}/A = (-1/2r)$, $\overline{H^2}/A = (-1/2r)^2$,  and attuned rod size for diffused layer is obtained by adding compact layer thickness to rod radius: $ r+r_H$. Here the diffuse layer starts after a distance $r+r_H$ from the center of each rod. The  average Gaussian curvature of the nanorods forest is calculated using the theorem of Gauss-Bonnet as  $\overline{K}/A = \int K dS /A = 4 \pi ( N_c - N_h)/A$. Using Eq.~\ref{E19} we obtained the mean capacitance for $N_r$ electrically connected forest of nanorods: $N_c =N_r$, $N_h = 0$  and $\overline{K}/A  =  (2/r l )$.
\begin{equation}
\langle c_S\rangle_{R} =  \frac{ c_H c_0  \left[  1 +   l_D/2(r+r_H) - l_{D}^{2}/8(r+r_H)^2 + l_{D}^{2}/l (r+r_H) \right] }{c_H + c_0 \left[  1 +   l_D/2(r+r_H) - l_{D}^{2}/8(r+r_H)^2 +  l_{D}^{2}/l (r+r_H)  \right] } \label{CapRod}
\end{equation} 
The relative mean square fluctuation in capacitance which arise due to surface fluctuations over rods is obtained using Eq.~\ref{rancap} as
\begin{equation}
\frac{\beta^2}{\langle c_G \rangle}  =  \frac{1}{c_{0}}\; \frac{\gamma^2\; l_{D}^{2}/4(r+r_H)^2}{\left[  1 +   l_D/2(r+r_H) - l_{D}^{2}/8(r+r_H)^2 +  l_{D}^{2}/l (r+r_H)  \right]^3} \label{ranrod}
\end{equation} 

\begin{figure}[ht]
\includegraphics[scale = 0.7]{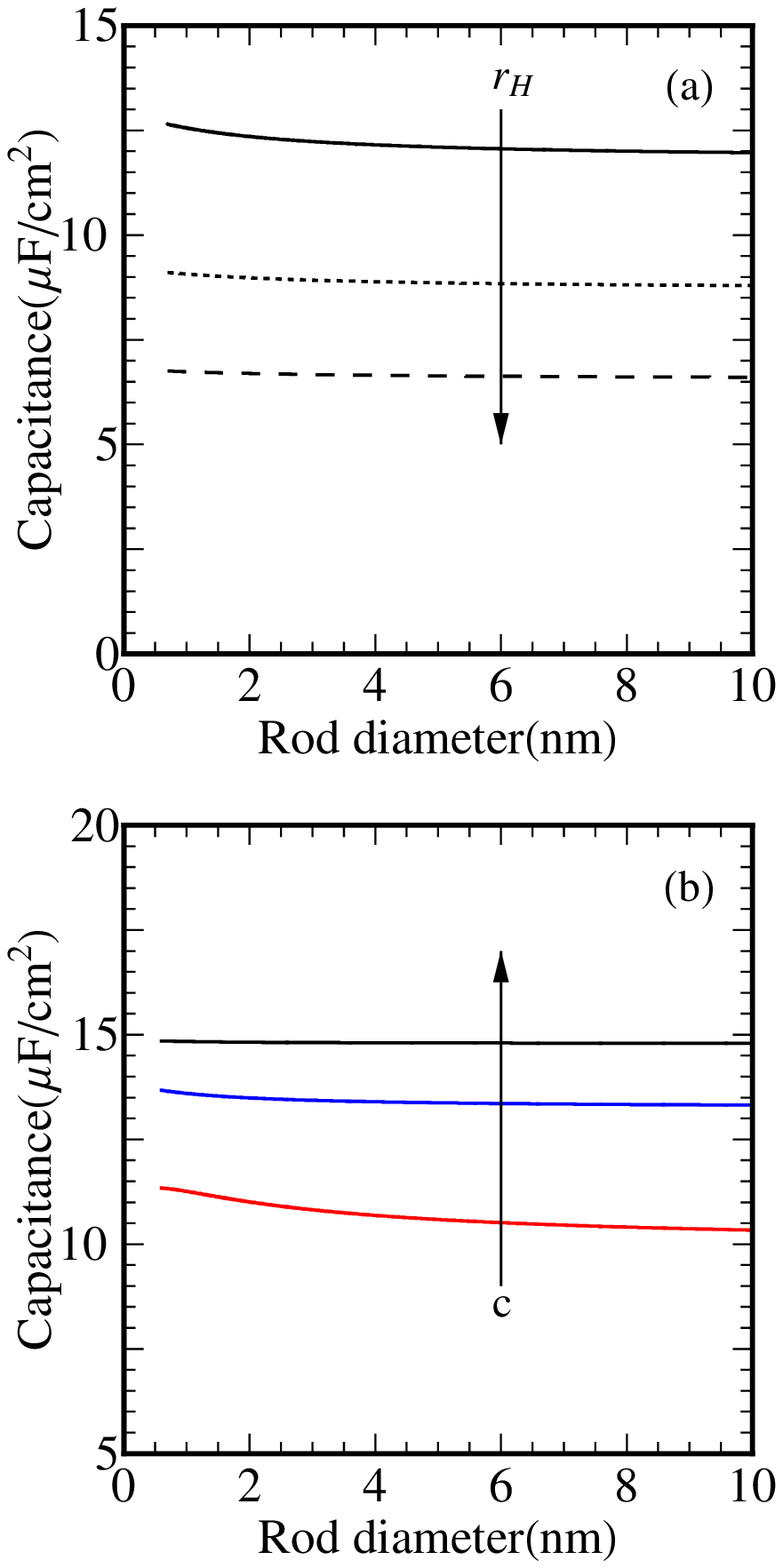}
\caption{Capacitance plots for an assembly of cylindrical rods with no morphological fluctuation ($\gamma=0$) obtained from Eq.~\ref{E21}. 
Fig. (a) shows the influence of the compact layer thickness ($r_H$), e.g. 0.34, 0.4 and 0.7 nm.
Fig. (b) shows the influence of the electrolyte molar concentration, e.g. 0.01, 0.1 and 1 M. All the plots are generated using $r_H$ = 0.34 nm. $\epsilon_H$ = 6 for   1:1 organic electrolyte  with $\epsilon$ = 38  at 298 K.}
\end{figure}

Fig. 5(a) shows the effect of HL thickness on the GCS capacitance density vs rod diameter plots in a forest of nanorods electrode where $\gamma = 0$. Plots are made using Eq.~\ref{CapRod}. As we increase the HL thickness, the value of capacitance decreases. The HL increases the attuned radii curvature of rods and hence resultant decrease in capacitance contributions from the diffuse layer. Fig. 5(b) shows the effect of diffuse layer thickness, which increases with decrease in the concentration of electrolyte, leads to decrease in capacitance. Thus for nano rods with the larger attuned radii of curvature, effect will largely be seen in the capacitance for  dilute solution. In case of concentrated solution, it simply increases the capacitance but no enhancement in capacitance is seen in microrods.

Fig. 5 (b) shows the effect of concentration on the capacitance of assembly of nanorods electrode using Eq.~\ref{CapRod} ( $\gamma = 0$ case). 
Unlike cylindrical pore geometry, the capacitance of rod geometry is enhanced with reduction of diameter. The overall capacitance is very much like the Helmholtz capacitance, but it is enhanced mainly for small rod diameter.  Increase in concentration results in  higher capacitance. We compare the capacitance obtained in porous membrane electrode with nanorods with same surface area, but different (local) convexity can have qualitatively different capacitance dependence on size.  These are purely geometric effects of the surface curvatures. Thus, it can be concluded that membrane electrode with cylindrical nanopores  have different capacitance as compared to forest of cylindrical nanorods although both  have same geometrical surface area. We see from the plots that the curvature strongly affects the diffuse layer capacitance.   Similar studies by  Compton and coworkers \cite{Dickinson, Henstridge} has shown that for hemispherical and nanotubes electrodes, there is enhanced diffuse double layer capacitance due to increase in curvature.  In fact, our analytical model calculations also show such an increase in diffuse double layer capacitance. But the introduction of compact (Helmholtz) layer capacitance in series with the diffuse layer capacitance reduces the total  capacitance. Here overall capacitance is controlled by smaller of the two. It is important to note that the curvature effect is very much reduced in case of  nanorods electrode unlike the diffuse layer capacitance without attuned radii correction.  Furthermore, as we decrease the concentration and hence increase in the Gouy length, we see the curvature effect to capacitance reduced. The diffuse layer is  extended in  low concentration regime and  may require accounting for overlapping of  EDL \cite{Farina, Hou}. 

\begin{figure}[ht]
\vskip0.5in
\includegraphics[scale = 0.5]{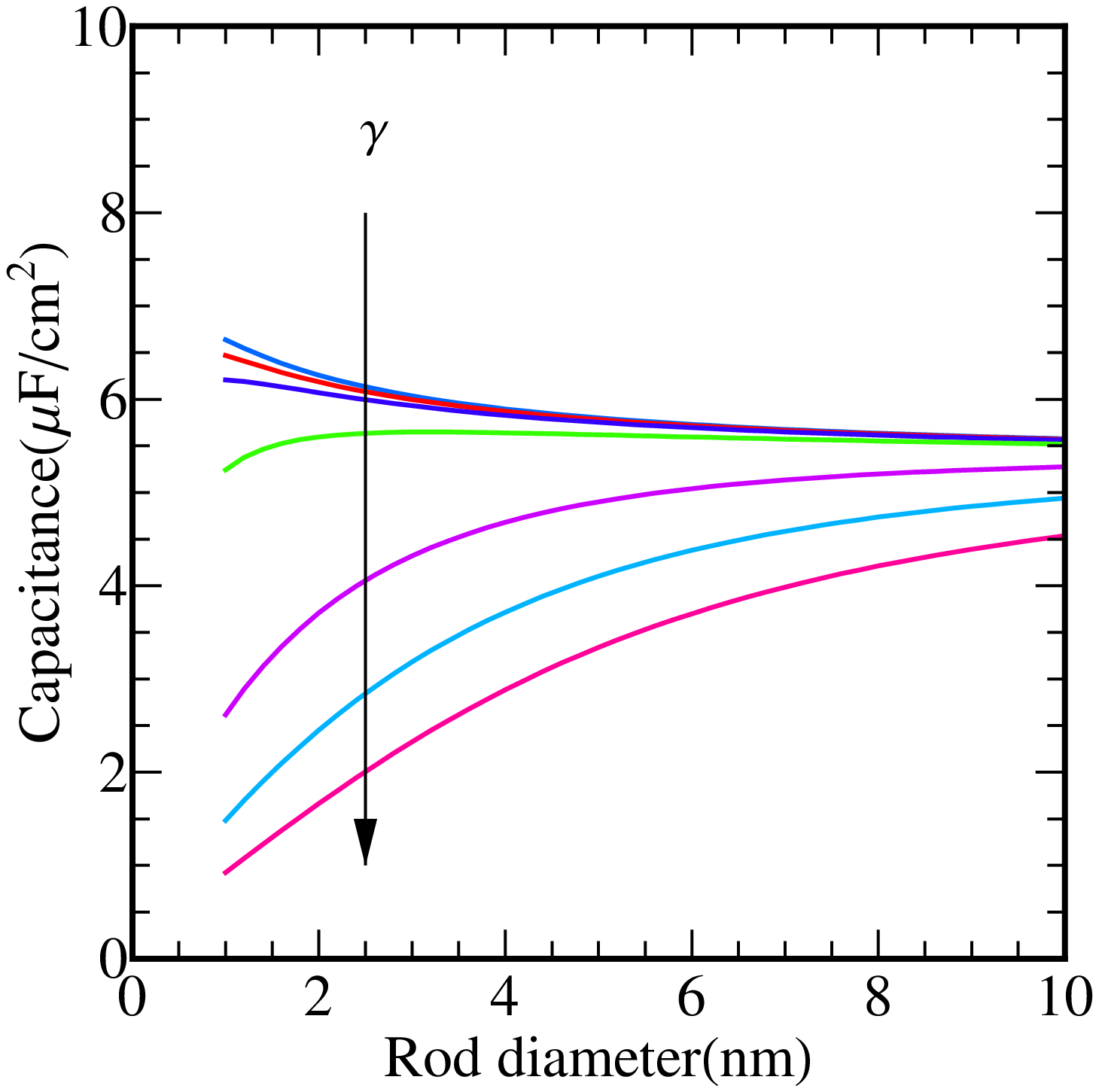}
\caption{Effect of morphological fluctuations on the capacitance of a nanorods forest electrode. Here fluctuation parameter is varied as: $\gamma$ = 0.3, 0.6, 0.9, 2, 4, 6, 8. All the plots are generated using $r_H$ = 0.25 nm, $\epsilon$ = 6, for 0.05 M $H_2 SO_4$ with $\epsilon$ = 78.6 at 298 K.}
\end{figure}

Fig. 6 shows the effect of morphological fluctuations on the capacitance of forest of nanorods. Increasing the curvature fluctuation parameter $(\gamma)$. The capacitance behaviour of  nanorods changes qualitatively from smooth nanorods capacitance behavior to nanoporous capacitance behaviour. For $\gamma$ = 2 , the curve nearly approaches  a horizontal line of a planar electrode, and effectively no curvature effects are seen. Curves show lowering in capacitance with an increase in the value $\gamma$ and curves appear similar to porous electrode capacitance with a gradual fall in smaller rod thickness.  The physical reason is, for an assembly of nanorods electrode with large curvature fluctuations on the nanorods  results in formation of pits or pore like structure which made the capacitive response  to behave like a porous system.

\section{Comparison of Theoretical Results with Experimental Data}

Fig. 7 shows the theoretical and experimental capacitance behavior of organic electrolyte TEABF$_4$ in CH$_3$CN   in different pore size regimes and concentrations. Data were taken from references\cite{Simon, Chmiola, HuangAC}. Dielectric constant of CH$_3$CN is taken to be 38 at 298K, bare ion diameter of TEA$^+$ is 0.68 nm, BF$_4$ $^-$ is 0.46 nm and solvent diameter is 0.34 nm\cite{Ue1994}. We use in our calculations for the  compact layer dielectric constant $\epsilon_H$ = 5.3 and the bare ion radius of TEA$^+$ as the compact layer thickness $r_H$ = 0.34 nm.
The plots are obtained using Eq.~\ref{E19} for smooth pores and Eq.~\ref{cap} in combination with Eqs.~\ref{E21} and \ref{ranpore}. We have mentioned earlier that we use bare ion size for the compact layer thickness and a reasonable value for the compact layer dielectric constant to evaluate four theoretical curves, as an illustration, of overall behavior of capacitance predicted by the theory and compared them with capacitance behavior in some of the recent data.
The theoretical plots for capacitance clearly indicate the minimum in an organic electrolyte. Black line is for 1.5 M electrolyte and electrode without morphological disorder, i.e. $\gamma = 0$ and black broken  line are for same concentration electrolyte but the electrode has some morphological disorder, i.e. $\gamma = 2$. Similarly, two lower curves have the electrode with same morphological disorder parameter $\gamma =6$ but differ in electrolyte concentration.  Two electrolyte concentrations 1.4 M and 1M, are compared with experimental data to illustrate the point that capacitance is dependent on electrolyte concentration. TEABF$_4$ in CH$_3$CN at 1.4 M have higher capacitance than 1 M TEABF$_4$ in CH$_3$CN for porous electrode with $\gamma = 6$. In agreement with the theoretical prediction, the effect of morphological fluctuation is seen upto 10 nm in the experimental data.

\begin{figure}[ht]
\includegraphics[scale = 1]{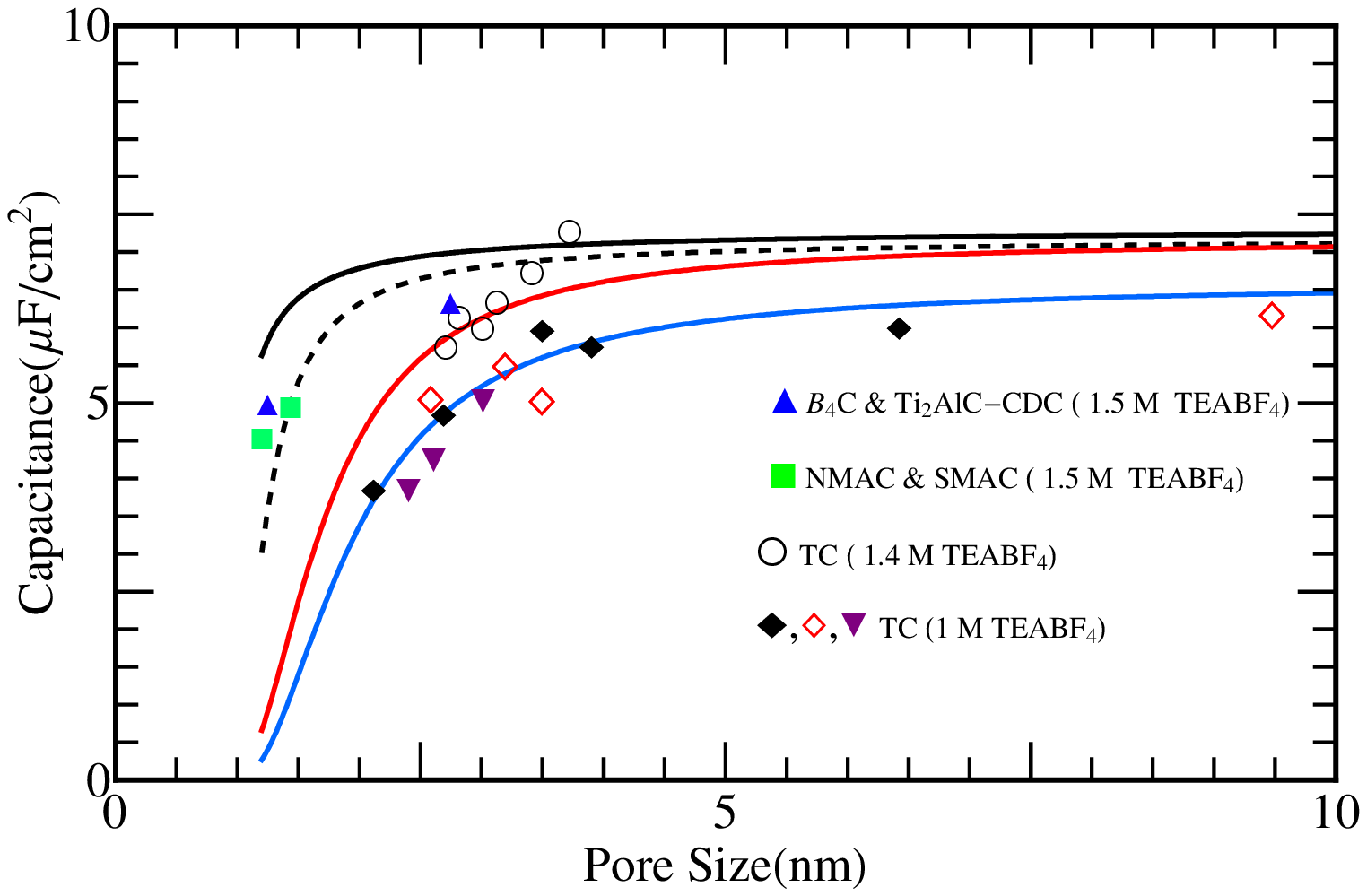}
\caption{Theoretical capacitance curves for a porous membrane electrode model of the electrode with and without morphological disorder.  Experimental data points are from ref \cite{HuangAC, Simon, Chmiola}. (a) Black line is for 1.5 M tetraethylammonium tetrafluoroborate (TEABF$_{4}$) organic electrolyte in acetonitrile (ACN)  with $\epsilon$ = 38  at 298 K, $\epsilon_H$ = 5.3, $r_H$ = 0.34nm, $\gamma$ =0.  (b) Broken black line is the same case as in (a) but with different $\gamma$ = 2. (c) Red line is for 1.4 M concentration and $\gamma$ = 6 while the rests of parameters are same as in (a). (d) Blue line for 1 M concentration and rest of system parameters are same as in (c).}
\end{figure}

Experimental comparison can further improve with  inclusion of the contribution from electronic space charge capacitance of the solid \cite{Gerischer85,Gerischer87,Kotz05}, as its contribution will be important in understanding the anomalous rise region and its maximum, but we have ignored this in the present calculations.

\section{Summary and Conclusions}

In summary, a model for EDL capacitance in curved nanostructured electrode is proposed within the frame work of classical (linearized) Gouy-Chapman-Stern theory. Our analytical theory for electric double layer capacitance of nanostructured and porous electrodes of arbitrary 
geometry  is applicable to the wider class of electrode morphology and electrolyte systems. Theory for arbitrary electrode geometry is developed using the ``multiple scattering formalism" in surface curvature, which includes geometrical as well as topological information about the surface. The region near the electrode is divided into a strong electric field region- the compact layer and the region beyond the compact layer with a relatively weak electric field region- the diffuse layer (assumed to obey linearized Gouy-Chapman equation). The capacitance  in the diffuse layer is expressed as a function through convergent expansion in power of the ratio  of Debye length to the principal radii of curvatures of the interface. The contribution of compact layer and its size correction in attuned geometric size in diffuse layer has a strong effect on the local capacitance. This is introduced through the capacitance in series behavior of local compact layer and local diffuse layer contributions. Theory is further extended for realistic electrodes with  a ubiquitous morphological disorder/roughness which is characterized through variance in mean curvatures. Our theory is general enough to handle arbitrary morphology of an electrode, but we illustrate its significance  for the simple system of forest of cylindrical nanopillars and nanoporous membrane with added complexity of morphological disorder. Theory allows us to analyze the effect of concentration, compact layer thickness, curvatures and its fluctuations along the pore.  

This paper further establishes for the capacitance of nanostructured electrodes the following points. The theory predicts a minimum value of the capacitance at minimum allowed mean pore size $2r_H+ l_D$ and a gentler rise towards larger pore sizes with a limiting capacitance plateau (at planar) as observed in experiments. The capacitances for nanoporous membrane electrode and nano-forest  electrode, with same geometric area,  have qualitatively different capacitance dependence on sizes. Hence, the overall  convexity of electrode surface has a profound effect on the ionic capacitance. The curvature fluctuations in nanostructured electrodes have profound influence on the capacitance in the mesoscopic region. The  small curvature fluctuations along the contour of pore or intrapore geometric fluctuations affect the  rise of capacitance which extends upto 2 nm size.  Hence, weak intrapore fluctuations slightly enhance the pore size dependence of capacitance. The effect of large curvature fluctuations  manifest themselves through prolonged dependence of capacitance on pore sizes and it persist in up-to nearly 10 nm in mesopore regime. Surprisingly, the large fluctuations in nano-rods forest electrode results in qualitatively different dependence  on rod thickness.

 Finally, 
our model is a step forward to develop a general theory of capacitance in complex interfacial systems. 
 This theory can be further extended by accounting quantum-mechanical contributions from electronic space charge in the electrode as well as more details in description of ion-ion correlations of EDL. 
  These extensions of the theory  may help to understand, the yet unsolved anomalous increase in capacitance in pore size less than 1 nm so called the   ``Gogotsi-Simon effect".
 Analogous problems of finding effect of  non-ideality of electrode surface, compact layer disorder and dynamic frequency-dependent response of nanostructured and porous electrodes will be reported elsewhere. \\ \\

\noindent{\bf Acknowledgement}
M. B. S. acknowledges the scholarship provided by UGC, New Delhi for  the Research Fellowship in Science for Meritorious Students. R.K. is grateful to University of Delhi and DST for purse grant.

\pagebreak
{\center \bf Appendix}\vskip.1in
\begin{appendix}
\setcounter{equation}{0}
\renewcommand{\theequation}{A. \arabic{equation}}

The Debye-H\"{u}ckel or linearized Gouy-Chapman equation for the potential relative to the bulk solution $\phi(r)$ is:
\begin{equation}
(\nabla^2-\kappa^2)\phi(r) = 0. \label{GCEA}
\end{equation}

For a domain $D$ bounded by a conducting surface $S$ which is held at constant potential $\phi^+|_S = \phi_0$ and far away from surface , $viz$. $\phi(r\longrightarrow\infty) = 
0$. The Green's function (GF) $G$ for a linearized  PB Eq.~\ref{GCEA} satisfies
\begin{equation} 
(\nabla^2-\kappa^2)G(r,r') = -\delta^{3}(r-r') \label{E3}
\end{equation}
with the homogeneous boundary conditions at surface $S$ and far away from surface ($r\longrightarrow\infty)$, $viz$. $G|_S = 
0$. The solution of Eq.~\ref{E3} for the potential relative to the bulk solution, $\phi(r)$ can be formally written as \cite{Duplantier90}
\begin{eqnarray}
\phi(r)&=&\phi_0\int_S dS_\beta \frac{\partial G(r, \beta^{+} )}{\partial n_\beta}\nonumber\\
&=&-\phi_0\int_V d^3r'\nabla^2 G(r, r' )\nonumber\\
&=&\phi_0\left(1-\kappa^2\int_V d^3r' G(r, r' )\right) \label{E5}
\end{eqnarray}

We seek to express $G$ in terms of $G_0(r,r') = ({1}/{4\pi|r-r'|})e^{-\kappa|r-r'|}$, the free space Green's function satisfying  Eq.~\ref{E3} for the entire infinite domain. The Green's function $G$ is expressed as \cite{Duplantier90},
\begin{eqnarray}
{ G}(r, r^\prime) &=& { G}_0 (r, r') - 2 \int { G}_0 (r, \alpha)
\frac{\partial { G}_0(\alpha , r^\prime)}{\partial n_\alpha} d S_\alpha \nonumber\\
&&+ 2^2 \int {G}_0 (r,\alpha)\frac{\partial  { G}_0(\alpha ,
\beta)}{\partial n_\alpha} \frac{\partial { G}_0(\beta , r^\prime)}{\partial n_\beta} d
S_\alpha d S_\beta 
\nonumber\\
&&+ \cdots  \label{E6}
\end{eqnarray} 
Using Gauss's law and the fundamental singularity at the boundary \cite{Balian,Duplantier90}, the differential capacitance density 
is 
\begin{eqnarray} 
 c(\alpha) & = & \frac{\epsilon \kappa^2}{4\pi} \int_V d^3 r' \frac{\partial G(\alpha^{+} ,r')}{\partial n_\alpha} \nonumber\\ 
& = & \frac {\epsilon \kappa^{2}}{4 \pi}  {\int_V d^3 r' }\left[  2 \frac{\partial G_0(\alpha ,r')}{\partial n_\alpha} \right.
\nonumber \\ 
&& \left. - 2^2\int \frac{\partial G_0(\alpha,\beta)}{\partial n_\alpha}\frac{\partial G_0(\beta,r')}{\partial n_\beta} dS_\beta \right. 
\nonumber\\ 
&& \left. + 2^{3} \int \frac {\partial G_0(\alpha , \beta)}{\partial n_\alpha} \frac{\partial G_0(\beta, \gamma)}{\partial n_\beta} \frac {\partial G_0(\gamma, r')}{\partial n_\gamma} dS_\beta dS_\gamma \right.
\nonumber\\
&& \left. -  \cdots \right] \label{E7}    
\end{eqnarray}  
The terms in Eq.~\ref{E7} can be looked upon as one-, two- and three-scattering terms as
\begin{equation}
\Sigma_1 = 2\int d^3 r'\frac{\partial G_0(\alpha ,r')}{\partial n_\alpha} \label{E8}
\end{equation}
\begin{equation}
\Sigma_2 = -2^2\int \frac{\partial G_0(\alpha, \beta )}{\partial n_\alpha}\frac{\partial G_0(\beta, r' )}{\partial n_\beta} dS_\beta d^3 r' \label{E9}
\end{equation}
\begin{equation}
\Sigma_3 = 2^3\int  \frac{\partial G_0(\alpha,\beta )}{\partial n_\alpha} \frac{\partial G_0(\beta, \gamma )}{\partial n_\beta} \frac{\partial G_0(\gamma, r')}{\partial n_\gamma} dS_\beta dS_\gamma d^3 r' \label{E10}
\end{equation}
Eq.~\ref{E8} may be expanded through the local surface coordinates $\alpha $, $\beta$, $\gamma$. For a weakly curved 
surface where Debye-H\"{u}ckel screening length is much smaller than the smallest scale of curvature the scattering kernel 
${\partial G_0(\beta^+,\alpha)} \slash {\partial n_\alpha}$ is expressed through a local coordinate system 
\cite{Balian,Duplantier90} with the z-axis parallel to inward normal vector $n_\alpha$ and a tangent plane on which projection is 
made. The local equation of surface, S in terms of curvature radii $R_1(\alpha)$ and $R_2(\alpha)$ : 
\begin{equation}
z_\alpha =({1}\slash{2})\left({x^2}\slash{R_1(\alpha)}+{y^2}\slash{R_2(\alpha)}\right) +\cdots  \label{E11}
\end{equation}
is introduced to Eq.~\ref{E8} through surface area element $ dS_\beta = \sqrt{g} dxdy $ where $ g = 1+ (\nabla 
z_\alpha(x,y))^{2}$. Using ${\partial}\slash{\partial n_\alpha}\equiv - {\partial}\slash{\partial z}$ the kernel ${\partial 
G_0(\alpha ,\beta)} \slash {\partial n_\alpha}$ under planar approximation\cite{Balian} reads  to first order as $({\partial 
G_0(\alpha ,\beta )}/{\partial n_\alpha}) = - ({z}\slash{\rho}) ({\partial G_0(\rho)}\slash{\partial \rho})$, where $\rho = |
\alpha'-\beta'| = (x^2+y^2)^{1\slash2}$ is distance in tangent plane , $G_0 = exp(-\rho)\slash 4 \pi \rho $ the Green's function 
in tangent plane. Now we can rewrite the one-scattering integral Eq.~\ref{E8} as
\begin{equation}%
\Sigma_1 = 2\int dx dy \left[ G_0(\rho) + \frac{1}{2}\frac{z^2}{\rho}\frac{\partial G_0}{\partial\rho}\right]  \label{E12}
\end{equation} 
which is further simplified using the angular averages ($<.>$)
\begin{equation}%
\frac{1}{\rho^2}\left\langle z_\alpha\right\rangle  =   \frac{1}{2}\int_0^{2\pi} d\theta \left( \frac{cos^2\theta}{R_1(\alpha)}+\frac{sin^2\theta}{R_2(\alpha)}\right) = \frac{\pi}{2}\left( \frac{1}{R_1(\alpha)}+\frac{1}{R_2(\alpha)}\right) \label{E13}
\end{equation}
\begin{eqnarray}%
\frac{1}{\rho^4}\left\langle z_\alpha^2\right\rangle & = &  \frac{1}{4}\int_0^{2\pi} d\theta \left( \frac{cos^2\theta}{R_1(\alpha)}+\frac{sin^2\theta}{R_2(\alpha)}\right)^2
\nonumber\\
& = & \frac{3 \pi}{8}\left( \frac{1}{R_1(\alpha){^2}}+\frac{1}{R_2(\alpha){^2}}\right) + \frac{\pi}{4 R_1(\alpha) R_2(\alpha)} \label{E14} 
\end{eqnarray}%
Substituting Eq.~\ref{E13} to \ref{E12} and  integrating over $\rho$ we finally get the one- scattering term as
\begin{equation}%
\Sigma_1 =  \frac{1}{\kappa}- \frac{1}{\kappa^3}\left( \frac{3}{2 R(\alpha)^2} -\frac{1}{2 R_1(\alpha) R_2(\alpha) }\right)+{\it O}\left(\frac{1}{R^3 }\right) \label{E15}
\end{equation}
where $H_\alpha=1/R(\alpha) = (1/2)(1/R_1(\alpha)+ 1/R_2(\alpha))$ and $K_\alpha=1/R_1(\alpha)R_2(\alpha)$. Similarly on iteration in Eq.~\ref{E9} and Eq.~\ref{E10}  two- and third-scattering terms are obtained as 
\begin{equation}%
\Sigma_2 =  - \frac{1}{\kappa^2}\frac{1}{R(\alpha) } + {\it O}\left(\frac{1}{R^3}\right),\; \Sigma_3  = \frac{1}{\kappa^3}\frac{1}{R(\alpha)^2 } + {\it O}\left(\frac{1}{R^3}\right) \label{E16}
\end{equation} 
Now substituting Eq.~\ref{E15} and Eq.~\ref{E16} to Eq.~\ref{E7}, after simplification the capacitance density at position $\alpha$ is 
\begin{equation}%
c (H_{\alpha},K_{\alpha}) = \frac{\epsilon \kappa}{4\pi}\left[  1- \frac{1}{\kappa}H_{\alpha} - \frac{1}{2\kappa^2}(H^2_{\alpha}-K_{\alpha}) + \cdots\right]   \label{E17B}
\end{equation}

\noindent

\end{appendix}

\end{document}